\documentclass[12pt]{article}
\usepackage{amsfonts}
\usepackage{amsmath}
\usepackage{amssymb}
\usepackage{xcolor}
\usepackage{color}
\usepackage[english]{babel}
\usepackage{graphics}
\usepackage{graphicx}
\usepackage{float}
\begin{document}
\title{
Survival amplitude, instantaneous energy \\ and decay rate of an unstable system:\\
Analytical results}
\author{\hfill \\K. Raczy\'{n}ska\footnote{e--mail: yaraczynska@gmail.com},  K. Urbanowski\footnote {e--mail:
K.Urbanowski@if.uz.zgora.pl}\\
University of Zielona G\'{o}ra, Institute of Physics, \\
ul. Prof. Z. Szafrana 4a, 65--516 Zielona G\'{o}ra, Poland.}
\maketitle
\vspace*{-24pt}
\begin{abstract}
We consider a model of a unstable state defined by the truncated Breit-Wigner energy density distribution function.
An analytical form of the survival amplitude $a(t)$ of the state considered is found. Our attention is focused
on the  late time properties of $a(t)$  and on effects generated by the non--exponential behavior of this amplitude
in the late time region:  In 1957 Khalfin proved that this  amplitude
tends to zero as $t$ goes to the infinity more slowly than any exponential function of $t$.  This
effect can be described using  a time-dependent decay rate $\gamma(t)$ and
then the Khalfin result means that this $\gamma(t)$ is not a constant but at late times  it tends to zero as $t$ goes to the infinity. It
appears that the energy $E(t)$ of the unstable state behaves similarly:
It tends to the minimal energy $E_{min}$ of the system as $t \to \infty$.
Within the model considered we find  two first leading  time dependent elements  of late time asymptotic expansions of
$E(t)$ and $\gamma (t)$. We discuss also possible implications of such a late time asymptotic properties of $E(t)$ and $\gamma (t)$
and cases where these properties may manifest themselves.
\end{abstract}

\section{Introduction}

Attempts to describe   time evolution of unstable states within the Quantum Mechanics were made practically from times when
this theory started to be born. The most  known result from these times is the Weisskopf--Wigner theory od spontaneous emission \cite{WW}.
Considering the excited atomic levels and applying the Shr\"{o}dinger equation to describe the time evolution Weisskopf and Wigner
found that to a good approximation the non--decay probability of the exited levels is a decreasing function of time having an exponential form \cite{WW}.
Further theoretical studies of the quantum decay process showed that basic principles of the quantum theory
does not allow it to be described by an exponential decay law at very late times
\cite{khalfin,fonda} and at initial stage of the decay process (see \cite{fonda} and references therein). Theoretical
analysis shows that at late times the survival probability (i. e. the decay law) should tends to zero as
$t \to \infty$ much more slowly than any exponential function of time and that as function of time it has the inverse power--like form
at this  regime of time \cite{khalfin,fonda}.
There was many unsuccessful attempts to verify experimentally predicted deviations from the exponential
form of the decay law at late times regime (see eg. \cite{norman}). The first experimental evidence of these
deviations  at long time regime was reported in \cite{rothe}.
Rothe and his group preparing their experiment
 used successfully conclusions resulting from theoretical studies of models of unstable states
and conditions leading to the non--exponential behavior
of the survival amplitude (see, e.g., \cite{seke} -- \cite{jiitoh}). The result reported by Rhote's group gives rise to another important problem:
 If (and how) the late time deviations from the exponential
decay  affect the energy of the unstable state and its decay
rate.
Theoretical studies of models of unstable states can bring us closer to understanding and explaining this problem.
This paper contains analysis of the quantum unstable system modeled by the Breit--Wigner energy
density distribution function. Studies of such models are known in the literature but usually these studies
were limited to the analysis of properties of the survival amplitude (see, eg. \cite{sluis} and \cite{ku-2008,ku-2009}).
The first leading late time terms of asymptotic series expansions for the energy $E(t)$ and decay
rate $\gamma(t)$ of the unstable state were found within such a model in \cite{ku-2008,ku-2009}. In this paper
we show how to find the 2nd or higher leading terms of the late time expressions for $E(t)$ and $\gamma(t)$.

The aim of this  paper is to find analytical expressions for the survival amplitude $a(t)$, the effective Hamiltonian
$h(t)$ governing the time evolution in the subspace of the unstable states considered and analytical late time
expressions for $a(t), h(t)$ and also of  the instantaneous energy $E(t)$ and decay rate $\gamma (t)$ with
the accuracy to the first two non--trivial leading elements of the asymptotic series expansions of these quantities within the model considered.

The paper is organized as follows. Section 2 contains a brief description of general properties of
evolving in time unstable states and basic definitions of quantities discussed in next Sections.
In Sect. 3 the model of an unstable state defined by the truncated Breit--Wigner energy density distribution
function is analyzed: There are found analytical expressions for  the survival amplitude $a(t)$, effective
Hamiltonian $h(t)$ as well as the late time asymptotic series expansions of $a(t)$, $h(t)$, $E(t)$ and $\gamma(t)$.
Section 4 contains graphical presentations of results of numerical calculations of
quantities  discussed in Section 3.
In Sec. 5 one finds a discussion and final remarks.

\section{Preliminaries}

Studying quantum  unstable systems one usually analyzes their decay law (that is in their survival probability), which contains
a main information about properties of such systems. If one knows that the system is in the initial unstable
state $|\phi\rangle \in {\cal H}$, (${\cal H}$ is
 the Hilbert space of states of the considered system), which was prepared at the initial instant $t_{0} =0$,
then one can calculate
its survival probability, ${\cal P}(t)$, of the unstable state $|\phi\rangle$ decaying
in vacuum, which equals
\begin{equation}
{\cal P}(t) = |a(t)|^{2}, \label{P(t)}
\end{equation}
where $a(t)$ is  the probability amplitude of finding the system at the
time $t$  in the initial unstable state $|\phi\rangle$,
\begin{equation}
a(t) = \langle \phi|\phi (t) \rangle . \label{a(t)}
\end{equation}
and $|\phi (t)\rangle$ is the solution of the Schr\"{o}dinger equation
for the initial condition  $|\phi (0) \rangle = |\phi\rangle$,
 \begin{equation}
i \hbar \frac{\partial}{\partial t} |\phi (t) \rangle = H |\phi (t)\rangle.  \label{Schrod}
\end{equation}
 Here $|\phi \rangle, |\phi (t)\rangle \in {\cal H}$,
  and
 $H$ denotes the total self--adjoint Hamiltonian for the system considered.
We assume that there exists a common inertial reference rest frame ${\cal O}_{0}$
for the observer and for the unstable system. So, ${\cal P}(t)$ is the
probability of finding the system at time $t$ in the rest reference frame ${\cal O}_{0}$ in the initial unstable state $|\phi\rangle$.

  An important property of the state
  $|\phi\rangle$ representing an unstable state
  is that the $|\phi\rangle$
 cannot be
 an eigenvector for $H$: Simply in such a case the eigenvalue equation $H|\phi\rangle = E_{\phi} |\phi\rangle$
 has no solutions.

An unstable state $|\phi\rangle$ can be modeled as a wave packets using solutions of the following eigenvalue equation
\begin{equation}
H|E\rangle = E|E\rangle, \label{H-m;0}
\;\;\; E\in \sigma_{c}(H),
\end{equation}
where $\sigma_{c}(H)$ denotes a continuum spectrum of $H$.
Eigenvectors $|E\rangle$ are normalized as usual:
\begin{equation}
\langle E|E'\rangle = \delta(E- E'). \label{d-m-m'}
\end{equation}
Using vectors $|E\rangle$ we can model an unstable state
as the  following wave--packet
\begin{eqnarray}
 |\phi\rangle \equiv |\phi\rangle
= \int_{E_{min}}^{\infty}\,c(E)\, |E\rangle\,dE,\label{phi}
\end{eqnarray}
where
expansion coefficients $c(E)$ are functions of the energy $E$ and
$E_{min}$ is the lower bound of the spectrum $\sigma_{c}(H)$ of $H$.
The state $|\phi\rangle$  is normalized $\langle \phi|\phi\rangle = 1$, which means that
it has to be $\int_{E_{min}}^{\infty}|c(E)|^{2}\,dE = 1$.

Using  the definition of the survival amplitude $a(t)$, the expansion (\ref{phi}) and the
relation (\ref{H-m;0}) we can find  $a(t)$, which takes the following form within the
formalism considered,
\begin{equation}
a(t)  \equiv \int_{E_{min}}^{\infty} \omega(E)\;
e^{\textstyle{-\,i\,E\,t}}\,d{E},
\label{a-spec}
\end{equation}
where $\omega(E) \equiv  |c(E)|^{2} > 0$.

As it is seen from (\ref{a-spec}), the amplitude $a(t)$, and thus the decay law ${\cal P}(t)$ of the
unstable state $|\phi\rangle$, are completely determined by the
density of the energy distribution $\omega(E)$ for the system
in this state \cite{fock}  (see also: \cite{khalfin,fonda,nowakowski,muga,muga-1,calderon-2,giraldi1,giraldi2}.
Now if to apply Riemann--Lebesque lemma to (\ref{a-spec}) then the conclusion follows:
$a(t) \to 0$ as $t \to \infty$. It is because  the normalization
condition $a(0) \equiv \int_{Spec. (H)} \,
\omega (E)\,dm = 1$ ensures the absolute integrability of  $\omega (E)$. So it has to be
${\cal P}(t) \to 0$ in the case considered.
(It appears that this approach can be  also applied for  Quantum Field Theory models \cite{giacosa2,giacosa,goldberger}).

Now if to follow
Khalfin   \cite{khalfin} and to assume
that the spectrum of $H$ must be bounded
from below, $E_{min} > - \infty$,
and to use
the Paley--Wiener
Theorem  \cite{Paley} then one comes to the Khalfin's conclusion
that in the case of unstable
states there must be
$|a(t)| \; \geq \; A\,\exp\,[- b \,t^{q}]$,
for $|t| \rightarrow \infty$. Here $A > 0,\,b> 0$ and $ 0 < q < 1$.
This means that
the decay law ${\cal P}(t)$ of unstable
states decaying in the vacuum, (\ref{P(t)}), can not be described by
an exponential function of time $t$ if time $t$ is suitably long, $t
\rightarrow \infty$, and that for these lengths of time ${\cal
P}(t)$ tends to zero as $t \rightarrow \infty$  more slowly
than any exponential function of $t$.
Not so long ago
this
this effect was confirmed by Rothe and his group
in experiment described  in
\cite{rothe}.

It appears that an
information about
the decay law ${\cal P}_{\phi}(t)$ of the state $|\phi\rangle$, strictly speaking
about the decay rate $\gamma_{0}$ of this state, as well
as the energy $E_{0}$ of the system in this state
can be be extracted from $a(t)$.
One can do this
using the rigorous equation governing the time evolution in the subspace of unstable
states, ${\cal H}_{\parallel} \ni |\phi\rangle_{\parallel} \equiv |\phi \rangle$.
Such an
equation
can be derived using the
Schr\"{o}dinger equation (\ref{Schrod}) for the total state
space ${\cal H}$. Namely
starting from (\ref{Schrod}) one finds
that within the problem considered.
 \begin{equation}
i \hbar \frac{\partial}{\partial t}\langle\phi |\phi (t) \rangle = \langle \phi|H |\phi (t)\rangle.  \label{h||1}
\end{equation}
So taking into account (\ref{a(t)}) it can be said
that the amplitude $a(t)$ satisfies the following equation
\begin{equation}
i \hbar \frac{\partial a(t)}{\partial t} = h(t)\,a(t), \label{h||2}
\end{equation}
where
\begin{equation}
h(t) = \frac{\langle \phi|H |\phi (t)\rangle}{a(t)} \equiv \frac{\langle \phi|H |\phi (t)\rangle}{\langle \phi|\phi (t)\rangle}, \label{h(t)-eq}
\end{equation}
and $h(t)$ is the effective Hamiltonian governing the time evolution in the subspace of unstable states
${\cal H}_{\parallel}= P {\cal H}$, where $P$ is the projection operator.
The subspace ${\cal H}_{\parallel}$ can describe an one--component unstable subsystem and then  $P= |\phi\rangle \langle \phi|$
(see \cite{pra}, \cite{ku-1991}  and also \cite{ku-2008,ku-2009} and references therein) or
multi--component subsystem (like neutral kaons complex) and other the like (see: \cite{ku-1992,ku-1995a,ku-1998} and references one can find therein).
 The subspace ${\cal H} \ominus {\cal H}_{\parallel} = {\cal H}_{\perp} \equiv Q {\cal H}$ is the subspace of decay products. Here $Q = \mathbb{I} - P$.
 An equivalent formula for $h(t)$ has the following form \cite{ku-2008,ku-2009,pra}:
\begin{equation}
h(t) \equiv \frac{i\hbar}{a(t)}\,\frac{\partial a(t)}{\partial t}. \label{h(t)}
\end{equation}

 The effective Hamiltonian $h(t)$ is used when one starts
with the Schr\"{o}\-din\-ger equation
for the total state space ${\cal H}$ and looks for the rigorous evolution equation for
a distinguished subspace of states ${\cal H}_{||} \subset {\cal H}$
(see, eg. \cite{pra} --- \cite{ku-1993} and also \cite{giraldi1,giraldi2}).
In general  $h(t)$ is a complex function of time. In the case of ${\cal H}_{\parallel}$ of dimension two or more
the effective Hamiltonian governing the time evolution in such a subspace   is a non--hermitian
matrix $H_{\parallel}$ or a non-hermitian operator \cite{ku-1991,ku-1992,ku-1993}.
We have
\begin{equation}
h(t) = E(t) - \frac{i}{2} \gamma (t), \label{h-m+g}
\end{equation}
 and
 \begin{equation}
 E(t) = \Re\,[h(t)], \;\;\;\;\;
\gamma (t) = -\,2\,\Im\,[h(t)],\label{m(t)}
\end{equation}
are the instantaneous mass energy   $E(t)$ and the instantaneous decay rate,
$\gamma (t)$ (see \cite{pra} and \cite{ku-2008,ku-2009}).
Here $\Re\,(z)$ and $\Im\,(z)$ denote the real and imaginary parts
of $z$ respectively. Relations (\ref{h||2}),  (\ref{h(t)}) and (\ref{m(t)}) are very helpful
when the density $\omega (E)$ is given and one wants to find
the instantaneous energy $E(t)$ and decay rate $\gamma (t)$: In such a case inserting
 $\omega (E)$ into (\ref{a-spec}) one obtains the amplitude $a(t)$ and
then using (\ref{h(t)}) one finds the $h(t)$ and thus $E(t)$ and $\gamma (t)$.

In closing this Section we should pay attention to another problem:
The  vector $|\phi\rangle$ of the form (\ref{phi})  describing  a quantum unstable subsystem can not be an eigenvector
of the Hamiltonian $H$,
otherwise it would be that
${\cal P}(t) =|\langle \phi|\phi(t)\rangle|^{2} = $  $|\langle \phi|\exp\,[-\frac{i}{\hbar}tH]\phi\rangle|^{2} \equiv 1$ for
all times $t$. The fact that  this vector $|\phi\rangle$
is not the
eigenvector for $H$ means that the energy of the quantum unstable object is not defined. Simply
the energy  can not take the exact constant value in this state $|\phi\rangle$.
In such a case
quantum systems are characterized by the energy   distribution density $\omega (E)$ and
the average energy   $<E> = \int_{E_{min}}^{\infty}\,E\,\omega(E)\,dE$ or by the instantaneous
energy  $E(t)$  but not by the exact value of the energy.

\section{A unstable system defined by the truncated Breit--Wigner energy density distribution: analytical results}

In the large literature many quantum unstable systems are described within the Fock--Krylov
theory using Breit--Wigner energy density distribution function $\omega_{BW}(E)$.
The use of $\omega_{BW}(E)$ is convenient because it describes relatively well a large class
of unstable systems and allows to find analytical form of the survival amplitude $a(t)$
(see, eg. \cite{sluis,ku-2008,ku-2009} and other papers). It appears that for this energy
density distribution one can find also analytical form of $a(t)$ at very late times as well
as analytical asymptotic form of $h(t), \;E(t)$ and $\gamma (t)$ for such times.

\subsection{A survival amplitude}
Let us assume that ${Spec. (H)} = [E_{min}, \infty)$ and let us choose
$\omega (E)$ as follows
\begin{equation}
\omega (E) \equiv \omega_{BW}(E) \stackrel{\rm def}{=} \frac{N}{2\pi}\,  {\mathit\Theta} (E - E_{min}) \
\frac{\gamma_{0}}{(E-E_{0})^{2} +
(\frac{\gamma_{0}}{2})^{2}}, \label{omega-BW}
\end{equation}
where $N$ is a normalization constant and ${\mathit\Theta} (E)$ is the unit step
function: ${\mathit\Theta} (E) = 1$ for $E \geq 0$ and ${\mathit\Theta} (E) = 0$ for $E < 0$.
For such $\omega (E)$ using the integral representation of the survival amplitude (\ref{a-spec})
one finds
\begin{equation}
a(t) = \frac{N}{2\pi}  \int_{E_{min}}^{\infty}
 \frac{{\gamma_{0}}}{(E- E_{0})^{2}
+ (\frac{\gamma_{0}}{2})^{2}}\, e^{\textstyle{ -
\frac{i}{\hbar}E \,t}}\,dE, \label{a-BW}
\end{equation}
where
\begin{equation}
\frac{1}{N} = \frac{1}{2\pi} \int_{E_{min}}^{\infty}
 \frac{\gamma_{0}}{(E- E_{0})^{2}
+ (\frac{\gamma_{0}}{2})^{2}}\, dE. \label{N}
\end{equation}
Using dimensionless variables
\begin{equation}
\eta = \frac{E - E_{min}}{\gamma_{0}},\;\;\; \beta = \frac{E_{0} - E_{min}}{\gamma_{0}}\;\;{\rm and}\;\;
\tau = \frac{\gamma_{0} t}{\hbar} \stackrel{\rm def}{=} \tau(t),\label{tau}
\end{equation}
the integral (\ref{a-BW}) can be rewritten in the following form
\begin{eqnarray}
a(t) \equiv a(\tau(t)) &=&\frac{N}{2\pi}\, e^{\textstyle{-\frac{i}{\hbar} E_{min}t}}\,e^{\textstyle{ -
i \beta \tau}}\;  \int_{-\beta}^{\infty}
 \frac{1}{\eta^{2} + \frac{1}{4}}\, e^{\textstyle{ -i\eta\tau}}\,d\eta, \label{a-BW-Emin-s1}\\
&=& \frac{N}{2\pi}\, e^{\textstyle{-\frac{i}{\hbar} E_{min}t}}\,
e^{\textstyle{ - i \beta \tau}}\;\times \;I_{\beta}(\tau), \label{I(t)a}
\end{eqnarray}
where
\begin{equation}
I_{\beta}(\tau) \stackrel{\rm def}{=}\int_{-\beta}^{\infty}
 \frac{1}{\eta^{2}
+ \frac{1}{4}}\, e^{\textstyle{ -i\eta\tau}}\,d\eta. \label{I(t)}
\end{equation}
After some algebra one can express the function $I_{\beta}(t)$ defined by the
 relation (\ref{I(t)}) in terms of the integral--exponential functions $E_{1}(z)$,
\begin{eqnarray}
I_{\beta}(t) = I_{\beta}(\tau(t))&\equiv& 2\pi e^{\textstyle{-\frac{\tau}{2}}}\,\Big\{1 -  \frac{i}{2\pi}
 \Big[
e^{\textstyle{\tau}}\,
E_{1}\Big(-i(\beta
+ \frac{i}{2} )\tau \Big) + \nonumber\\
&&\;\;\;\;\;+ (-1) E_{1}\Big(- i(\beta -
\frac{i}{2})\tau\Big)\,\Big]\, \Big\}, \label{I(tau)}
\end{eqnarray}
wherein  $E_{1}(z)$ is defined according to formula 6.2.1 in
\cite{olver}. From (\ref{I(tau)}) and (\ref{I(t)a}) the following formula follows
\begin{eqnarray}
a(t) \equiv a(\tau (t)) &=&  N\,e^{\textstyle{- i (\beta\, - \,\frac{i}{2})\,\tau }} \times \nonumber \\
&&\times  \Big\{
\,e^{\textstyle{-\frac{i}{\hbar} E_{min}t}}\,
- \frac{i}{2\pi} \Big[
e^{\textstyle{\tau}}\,
E_{1}\Big(-i(\beta
+ \frac{i}{2} )\tau \Big) \nonumber\\
&&\;\;\;\;\;+ (-1) E_{1}\Big(- i(\beta -
\frac{i}{2})\tau\Big)\,\Big]\, \Big\}, \label{a-E(1)-tau}
\end{eqnarray}
or equivalently,
\begin{eqnarray}
a(t) &=& N\,e^{\textstyle{- \frac{i}{\hbar} (E_{0} -
i\frac{\gamma_{0}}{2})t}} \times \nonumber\\
&&\times \Big\{1 - \frac{i}{2\pi} \Big[
e^{\textstyle{\frac{\gamma_{0}t}{\hbar}}}\,
E_{1}\Big(-\frac{i}{\hbar}(E_{R}
+ \frac{i}{2} \gamma_{0})t\Big)  + \nonumber\\
&&\;\;\;\;+ (-1) E_{1}\Big(- \frac{i}{\hbar}(E_{R} -
\frac{i}{2} \gamma_{0})t\Big)\,\Big]\, \Big\}, \label{a-E(1)}
\end{eqnarray}
where $E_{R} = E_{0} - E_{min}$.

The results    (\ref{a-E(1)-tau}) and  (\ref{a-E(1)}) mean
in general $|a(t)|^{2}$ is not a pure exponential function of time that within the model considered. What is more it appears that
the survival amplitude $a(t)$ can not coincide with the canonical survival amplitude $a_{c}(t)$,
\[
a_{c}(t) \stackrel{\rm def}{=} A\,e^{\textstyle{ - \frac{i}{\hbar}(E_{0} - \frac{i}{2}\gamma_{0})t}},
\]
at any finite time interval. In order to see that
let us assume that there exist such a time interval $[t_{1}, t_{2}]$, where $t_{1} < t_{2}$, that
\[
a(t) = a_{c}(t),
\]
for all $t\in[t_{1},t_{2}]$. Then using (\ref{a-E(1)}) one finds that it should be in such a case:
\begin{eqnarray}
 A\,e^{\textstyle{ - \frac{i}{\hbar}(E_{0} - \frac{i}{2}\gamma_{0})t}}
&\equiv& N\,e^{\textstyle{- \frac{i}{\hbar} (E_{0} -
i\frac{\gamma_{0}}{2})t}} \times \nonumber\\
&&\times \Big\{1 - \frac{i}{2\pi} \Big[
e^{\textstyle{\frac{\gamma_{0}t}{\hbar}}}\,
E_{1}\Big(-\frac{i}{\hbar}(E_{R}
+ \frac{i}{2} \gamma_{0})t\Big) \nonumber\\
&&\;\;\;\;\;+ (-1) E_{1}\Big(- \frac{i}{\hbar}(E_{R} -
\frac{i}{2} \gamma_{0})t\Big)\,\Big]\, \Big\}, \label{a-ac}
\end{eqnarray}
This means that
\begin{eqnarray}
1 - \frac{A}{N} &=&
\frac{i}{2\pi} \Big[
e^{\textstyle{\frac{\gamma_{0}t}{\hbar}}}\,
E_{1}\Big(-\frac{i}{\hbar}(E_{R}
+ \frac{i}{2} \gamma_{0})t\Big) \nonumber\\
&&\;\;\;\;\;+ (-1) E_{1}\Big(- \frac{i}{\hbar}(E_{R} -
\frac{i}{2} \gamma_{0})t\Big)\,\Big], \label{a-ac-1}
\end{eqnarray}
or,
\begin{multline}
\chi (t) \stackrel{\rm def}{=}
e^{\textstyle{\frac{\gamma_{0}t}{\hbar}}}\,
E_{1}\Big(-\frac{i}{\hbar}(E_{R}
+ \frac{i}{2} \gamma_{0})t\Big) \\
- E_{1}\Big(- \frac{i}{\hbar}(E_{R} -
\frac{i}{2} \gamma_{0})t\Big)
= c \equiv const., \label{a-ac-2}
\end{multline}
for $t\in[t_{1},t_{2}]$.
So, there should be in the considered case,
\[
\frac{d \chi (t)}{dt} = 0,
\]
for $t\in[t_{1},t_{2}]$. Using (\ref{a-ac-2}) one finds that
\begin{equation}
\frac{d \chi (t)}{dt} \equiv \frac{\gamma_{0}}{\hbar} \,e^{\textstyle{\frac{\gamma_{0} t}{\hbar}}}\,
E_{1}\Big(-\frac{i}{\hbar}(E_{R} + \frac{i}{2} \gamma_{0})t \Big). \label{d-chi}
\end{equation}
From the last relation and from properties of integral--exponential function $E_{1}(z)$ one concludes
that the condition $\frac{\partial \chi(t) }{d t} = 0$ can be satisfied  for isolated values of time
$t$ at most. This means that there does not exist any time interval $[t_{1}, t_{2}]$ such that
 $\chi (t) =const$ for all $t\in[t_{1}, t_{2}]$, or that there does not exist any time interval
 $[t_{1}, t_{2}]$ and any $A\neq 0$ such that  for  $t\in[t_{1}, t_{2}]$ it could be
  ${\cal P}(t) \equiv |a(t)|^{2} = |a_{c}(t)|^{2} \equiv |A|^{2} \,\exp\,[-\gamma_{0}t]$ in the considered model defined by $\omega (E)=\omega_{BW}(E)$.

\subsection{Instantaneous energy $E(t)$ and decay rate $\gamma (t)$}

Not let us analyze properties of the instantaneous energy $E(t)$ and instantaneous decay
rate $\gamma (t)$ in the model considered. These quantities are defined using the effective Hamiltonian $h(t)$.
In order to find $h(t)$ we need
for the quantity  $i\,\hbar\, \frac{\partial a(t)}{\partial t}$ (see (\ref{h(t)})).
From  (\ref{I(t)a})
one finds that
\begin{eqnarray}
i \hbar \frac{\partial a(t)}{\partial t} &=& E_{0} \,a(t)\,
+ \,\gamma_{0}\,\frac{N}{2\pi}\,e^{\textstyle{-\frac{i}{\hbar} E_{0}t}}\,J_{\beta}(\tau(t)), \label{J-R-1}
\end{eqnarray}
where
\begin{equation}
J_{\beta}(\tau) = \int_{- \beta}^{\infty}\,\frac{x}{x^{2} + \frac{1}{4}}\,e^{\textstyle{-ix\tau}}\,dx, \label{J-R}
\end{equation}
or simply (see (\ref{I(t)})),
\begin{equation}
J_{\beta}(\tau) \equiv i\frac{\partial I_{\beta} (\tau)}{\partial \tau}. \label{J-R-eq}
\end{equation}
This last relation is very convenient when one tries to find an analytical expression for
$J_{\beta}(\tau) $: One just has  to find an analytical formula for $I_{\beta} (\tau)$ and then  use relation (\ref{J-R-eq}).

Now the use  (\ref{I(t)a}), (\ref{J-R-1}) and (\ref{h(t)}) leads to the conclusion that
\begin{equation}
h(t) = i \hbar \frac{1}{a(t)}\,\frac{\partial a(t)}{\partial t} = E_{0} + \gamma_{0}\,\frac{J_{\beta}(\tau(t))}{I_{\beta}(\tau(t))}, \label{h(t)-1}
\end{equation}
which means that
\begin{equation}
E(t) = \Re\,[h(t)] = E_{0}  + \gamma_{0}\,\Re\,\left[\frac{J_{\beta}(\tau(t))}{I_{\beta}(\tau(t))}\right], \label{E(t)-2}
\end{equation}
and
\begin{equation}
\gamma(t) = -2\,\Im[h(t)] = -2\,\gamma_{0}\,\Im\left[\frac{J_{\beta}(\tau(t))}{I_{\beta}(\tau(t))}\right]. \label{g(t)-a}
\end{equation}

In order to visualize properties of $E(t)$ it is convenient to use the following function:
\begin{equation}
\kappa (t) \stackrel{\rm def}{=}  \frac{E(t) - E_{min}}{E_{0} - E_{min}} . \label{kappa}
\end{equation}
Using (\ref{E(t)-2}) one finds that
\begin{equation}
E(t) - E_{min} = E_{0} - E_{min} + \gamma_{0}\,\Re\,\Big[\frac{J_{\beta}(\tau)}{I_{\beta}(\tau)}\Big], \label{E(t)-1}
\end{equation}
If to divide two sides of the above equation by $E_{0} - E_{min}$ then one obtains the function $\kappa (t)$ (see (\ref{kappa})) we are looking for:
\begin{equation}
\kappa (\tau(t)) = 1 + \frac{1}{\beta}\,\Re\,\Big[\frac{J_{\beta}(\tau(t))}{I_{\beta}(\tau(t))}\Big]. \label{kappa-1}
\end{equation}

An alternative analytical formula for $h(t)$ can be obtained using analytical
expressions for $a(t)$ and $\frac{\partial a(t)}{\partial t}$.
One can express the integral (\ref{J-R})  defining $J_{\beta}(\tau)$ in terms of the
exponential integral functions, which allows us to rewrite the formula
(\ref{J-R-1}) for $i\hbar \frac{\partial a(t)}{\partial t}$ as follows
\begin{eqnarray}
i \hbar \frac{\partial a(t)}{\partial t} &=& E_{0} \,a(t)\, +\, (-i)\,\gamma_{0}\,\frac{N}{2}\,e^{\textstyle{-\frac{i}{\hbar}(E_{0} - \frac{i}{2}\gamma_{0})t}}\,
\times \nonumber \\
&& \times \Big\{1\,
 +\, \frac{i}{2\pi}\Big[e^{\textstyle{\tau}}\,E_{1}[-i (\beta +\frac{i}{2})\tau]
 + E_{1}[-i(\beta - \frac{i}{2})\tau]\Big]\Big\}. \label{da-dt-1}
\end{eqnarray}
This relation together with (\ref{a-E(1)-tau}) gives
\begin{eqnarray}
h(t) &\equiv & E_{0} - \frac{i}{2}\,\gamma_{0}\,\times \nonumber \\
&& \times
\frac{1 + \frac{i}{2\pi} \Big[e^{\textstyle{\tau}}\,E_{1}[-i (\beta +\frac{i}{2})\tau]
 + E_{1}[-i(\beta - \frac{i}{2})\tau]\Big] }{1\,
- \frac{i}{2\pi} \Big[ e^{\textstyle{\tau}}\,
E_{1}\Big(-i(\beta
+ \frac{i}{2} )\tau \Big)
 -  E_{1}\Big(- i(\beta -
\frac{i}{2})\tau\Big)\,\Big]}. \label{h(t)-alt}
\end{eqnarray}
Taking the real part of $h(t)$ given by the last relation one obtains $E(t) \equiv \Re\,[h(t)]$
and then one can calculate $\kappa (t)$ and so on.

\subsection{Late time properties of $a(t),\,h(t)$ and $\gamma (t)$}

The late time asymptotic form of  $a(t)$ and $h(t)$ can be found using e.g. the formulae
(\ref{I(t)}),  (\ref{h(t)-1}) respectively. In order to do this for the beginning we should
find the asymptotic form of integrals $I_{\beta}(\tau)$ and $J_{\beta}(\tau)$ used in the
formula (\ref{h(t)-1}) for $h(t)$. These integrals are defined by the expressions (\ref{I(t)}) and (\ref{J-R}) respectively.
It is relatively simple to find asymptotic expressions $I_{\beta}(\tau)$
and $J_{\beta}(\tau)$ for $\tau \to \infty$ directly from (\ref{I(t)})
and (\ref{J-R}) using , e.g., the method of the integration by parts. We have for $\tau \to \infty$
\begin{multline}
I_{\beta}(\tau) \simeq \frac{i}{\tau}\,\frac{e^{\textstyle{i\beta \tau}}}{\beta^{2} + \frac{1}{4}}\,
\Big\{-1 \,+ \,\frac{2 \beta}{\beta^{2} + \frac{1}{4}}\,\frac{i}{\tau} \, + \,\frac{2}{\beta^{2} + \frac{1}{4}}\Big[ 1- \frac{4 \beta^{2}}{\beta^{2} + \frac{1}{4}}\Big]\,\Big(\frac{i}{\tau}\Big)^{2} \\
+\,\frac{24\beta}{(\beta^{2} + \frac{1}{4})^{2}}\Big[\frac{2\beta^{2}}{\beta^{2} + \frac{1}{4}}\,-\,1\Big]\,\Big(\frac{i}{\tau}\Big)^{3} \\
+\,\frac{24}{(\beta^{2} + \frac{1}{4})^{2}}\Big[-\,\frac{16\beta^{4}}{(\beta^{2} + \frac{1}{4})^{2}}\,+\,\frac{12\beta^{2}}{\beta^{2} + \frac{1}{4}}\,-\,1\Big]
\,\Big(\frac{i}{\tau}\Big)^{4}\; +\;\ldots \Big\}, \label{I-as}
\end{multline}
and
\begin{multline}
J_{\beta}(\tau) \simeq \frac{i}{\tau}\,\frac{e^{\textstyle{i\beta \tau}}}{\beta^{2} + \frac{1}{4}}
\Big\{\beta +
\Big[1 - \frac{2 \beta^{2}}{\beta^{2} + \frac{1}{4}}\Big]\,\frac{i}{\tau} + \frac{2\beta}{\beta^{2} + \frac{1}{4}}\Big[\frac{4 \beta^{2}}{\beta^{2} + \frac{1}{4}}\, -\, 3\Big]\,\Big(\frac{i}{\tau}\Big)^{2}\\
+\,\frac{6}{\beta^{2} + \frac{1}{4}}\Big[-\frac{8\beta^{4}}{(\beta^{2} + \frac{1}{4})^{2}}\,+\,\frac{8\beta^{2}}{\beta^{2} + \frac{1}{4}}\,-\,1  \Big]\,\Big(\frac{i}{\tau}\Big)^{3}\\
+\,\frac{24\beta}{(\beta^{2} + \frac{1}{4})^{2}}\Big[ \frac{16\beta^{4}}{(\beta^{2} + \frac{1}{4})^{2}}\,-\,\frac{20\beta^{2}}{\beta^{2} + \frac{1}{4}}\,+\,5 \Big]\,\Big(\frac{i}{\tau}\Big)^{4}\;
\ldots \Big\}. \label{J-as}
\end{multline}
These two last asymptotic expressions allow one to find for $\tau \to \infty$ the asymptotic
form of the ratio $\frac{J_{\beta}(\tau)}{I_{\beta}(\tau)}$ used in relations
(\ref{h(t)-1}), (\ref{E(t)-2}) and (\ref{E(t)-1}), which has much simpler form than
asymptotic expansions for $I_{\beta}(\tau)$ and $J_{\beta}(\tau)$.
In order to do this let us define
an auxiliary function $\phi (x)$,
\begin{equation}
\frac{J_{\beta}(\tau)}{I_{\beta}(\tau)} \equiv \phi (x) = \frac{\beta + a_{1}x + a_{2}x^{2} + a_{3}x^{3} + a_{4}x^{4} + \ldots}{-1 + b_{1}x +b_{2}x^{2} + b_{3}x^{3} + b_{4}x^{4} \ldots}.\label{phi2}
\end{equation}
where $x = i/\tau$. Now taking into account that at late times $1/\tau \ll 1$, which means that $|x| \ll 1$ at this times region and
expanding $\phi (x)$ given by (\ref{phi2}) in Taylor series around $x=0$ one finds after some algebra   that
\begin{eqnarray}
\phi (x) & \simeq & - \beta\, - \,x\,+\, \frac{2\beta}{\beta^{2} + \frac{1}{4}}\,x^{2}\,+ \nonumber \\
&& + \,\frac{1}{4}\; \frac{1 + 24 \beta - 28 \beta^{2} - 96 \beta^{3} + 64\beta^{4}}{(\beta^{2} + \frac{1}{4})^{3}}\,x^{3} \nonumber \\
&&  + \,\frac{1}{4}\;\frac{6 - 21 \beta + 48\beta^{2} - 64 \beta^{3} - 288 \beta^{4} + 464\beta^{5}}{(\beta^{2} + \frac{1}{4})^{4}}\,x^{4}\,+
\ldots \; , \label{phi3}
\end{eqnarray}
for $|x| \ll 1$. Hence,
there is for $\tau \to \infty$,
\begin{eqnarray}
\frac{J_{\beta}(\tau)}{I_{\beta}(\tau)} &\simeq & - \,\beta \;-\;\frac{i}{\tau}\;-\; \frac{2 \beta}{\beta^{2} + \frac{1}{4}}\,\frac{1}{\tau^{2}} \nonumber \\
&& -\,i\,\frac{1}{4}\; \frac{1 + 24 \beta - 28 \beta^{2} - 96 \beta^{3} + 64\beta^{4}}{(\beta^{2} + \frac{1}{4})^{3}}\,\frac{1}{\tau^{3}} \nonumber \\
&& +\,\frac{1}{4}\;\frac{6 - 21 \beta + 48\beta^{2} - 64 \beta^{3} - 288 \beta^{4} + 464\beta^{5}}{(\beta^{2} + \frac{1}{4})^{4}}\,\frac{1}{\tau^{4}} +
\,\ldots ,\label{J-I-as}
\end{eqnarray}
where $\tau$ and $\beta$ are defined by formulae (\ref{tau})   respectively.

Using relation (\ref{I(t)a}) one finds  from (\ref{I-as}) the late time asymptotic form $a_{lt}(t)$ of the survival amplitude $a(t)$. There is
\begin{eqnarray}
a_{lt}(t) &\stackrel{\rm def}{=}& {a(t)\vline}_{t \to \infty}
=  \frac{N}{2\pi}\,\frac{ e^{\textstyle{-\frac{i}{\hbar} E_{min}t}}}{\beta^{2} + \frac{1}{4}}\,\times\, \nonumber \\
&& \times\,
\Big\{
- i \, \frac{\hbar}{\gamma_{0}\,t} \, - \,\frac{2 \beta}{\beta^{2} + \frac{1}{4}} \, \Big(\frac{\hbar}{\gamma_{0}\,t} \Big)^{2}  \nonumber \\
&& -i \,\frac{2}{\beta^{2} + \frac{1}{4} } \Big[ 1- \frac{4 \beta^{2}}{\beta^{2} + \frac{1}{4}} \Big]\,\Big( \frac{\hbar}{\gamma_{0}\,t} \Big)^{3} \nonumber  \\
 && +\,\frac{24\beta}{(\beta^{2} + \frac{1}{4})^{2}}\Big[\frac{2\beta^{2}}{\beta^{2} + \frac{1}{4}}\,-\,1\Big]\,\Big(\frac{\hbar}{\gamma_{0}\,t}\Big)^{4}
\; +\;\ldots
\Big\}. \label{a-lt}
\end{eqnarray}

Starting from the asymptotic expression (\ref{J-I-as})   and using formula (\ref{h(t)-1}) one can
find the late time asymptotic form of $h(t)$ and thus of $E(t)$ and $\gamma (t)$ for model considered,
\begin{eqnarray}
{E(t)\vline}_{\,t \rightarrow \infty} &= &{\Re\,[h(t)] \vline}_{t \to \infty} \label{Re-h-as} \\
&\simeq& { E}_{\text{min}}\, -\,2\,
\frac{ { E}_{0}\,-\,{E}_{min}}{ \gamma_{0}^{2}\,(\beta^{2} + \frac{1}{4}) } \,
\left(\frac{\hbar}{t} \right)^{2}\nonumber \\
&& +\,\frac{1}{4}\,\frac{6 - 21 \beta + 48\beta^{2} - 64 \beta^{3} - 288 \beta^{4} + 464\beta^{5}}{\gamma_{0}^{3}\,(\beta^{2} + \frac{1}{4})^{4}}\,\left(\frac{\hbar}{t} \right)^{4}+\ldots,\nonumber
\end{eqnarray}
and,
\begin{eqnarray}
{{\it\gamma}(t)\vline}_{\,t \rightarrow \infty} &=& - 2 \Im\,[h(t)] \label{Im-h-as} \\
&\simeq & 2\,\frac{\hbar}{t} +\frac{1}{2}\, \frac{1 + 24 \beta - 28 \beta^{2} - 96 \beta^{3} + 64\beta^{4}}{\gamma_{0}^{2}\,(\beta^{2} + \frac{1}{4})^{3}}\,
\left(\frac{\hbar}{t}\right)^{3} +
\ldots \, . \nonumber
\end{eqnarray}
These three  last relations are valid for $t > T$, where $T$ denotes the cross--over time, i.e.
the time when canonical exponential and late time inverse power law contributions to the survival amplitude become comparable:
\begin{equation}
|a_{c}(t)|^{2} \simeq |a_{lt}(t)|^{2}. \label{ac=alt}
\end{equation}
The cross--over time $T$ is the solution of the last equation.

\section{Numerical results}
\begin{figure}[H]
\begin{center}
{\includegraphics[width=74mm]{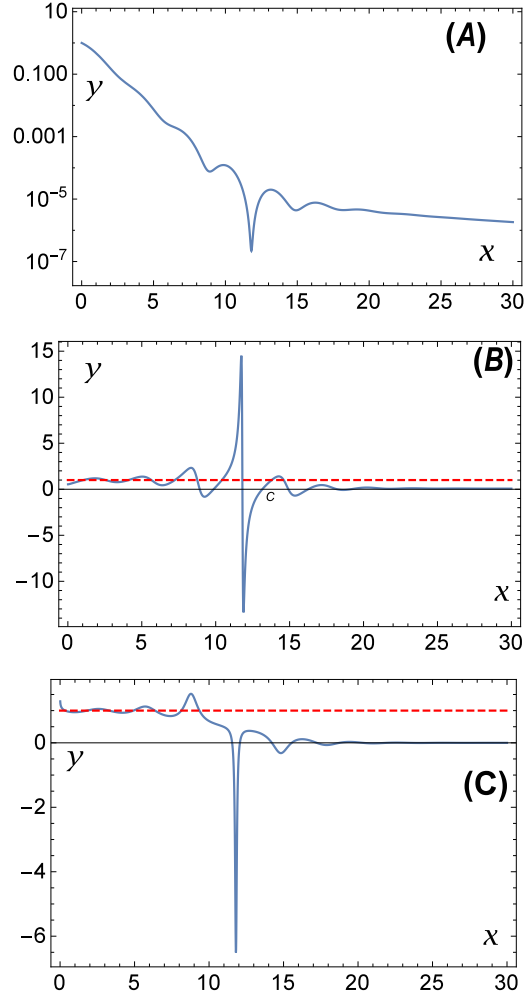}}
\caption{ $(A)$ The decay curve, $(B)$ The instantaneous decay rate, $(C)$ The instantaneous energy. Axes: $x$ --- in all panels time $t$ measured in lifetimes  $\tau$: $x = t/\tau$;  $(A)$ --- $y = {\cal P}(t/\tau)$, (The logarithmic scale); $(B)$ --- $y = \gamma (t/\tau)/ \gamma_{0}$; $(C)$ --- $\kappa (t/\tau)$, ($\kappa (t/\tau)$ is defined by formula (\ref{kappa})). The case  $\beta=2$.}
  \label{f1}
  \end{center}
  \end{figure}

\begin{figure}[H]
\begin{center} %{center}
{\includegraphics[width=78mm]{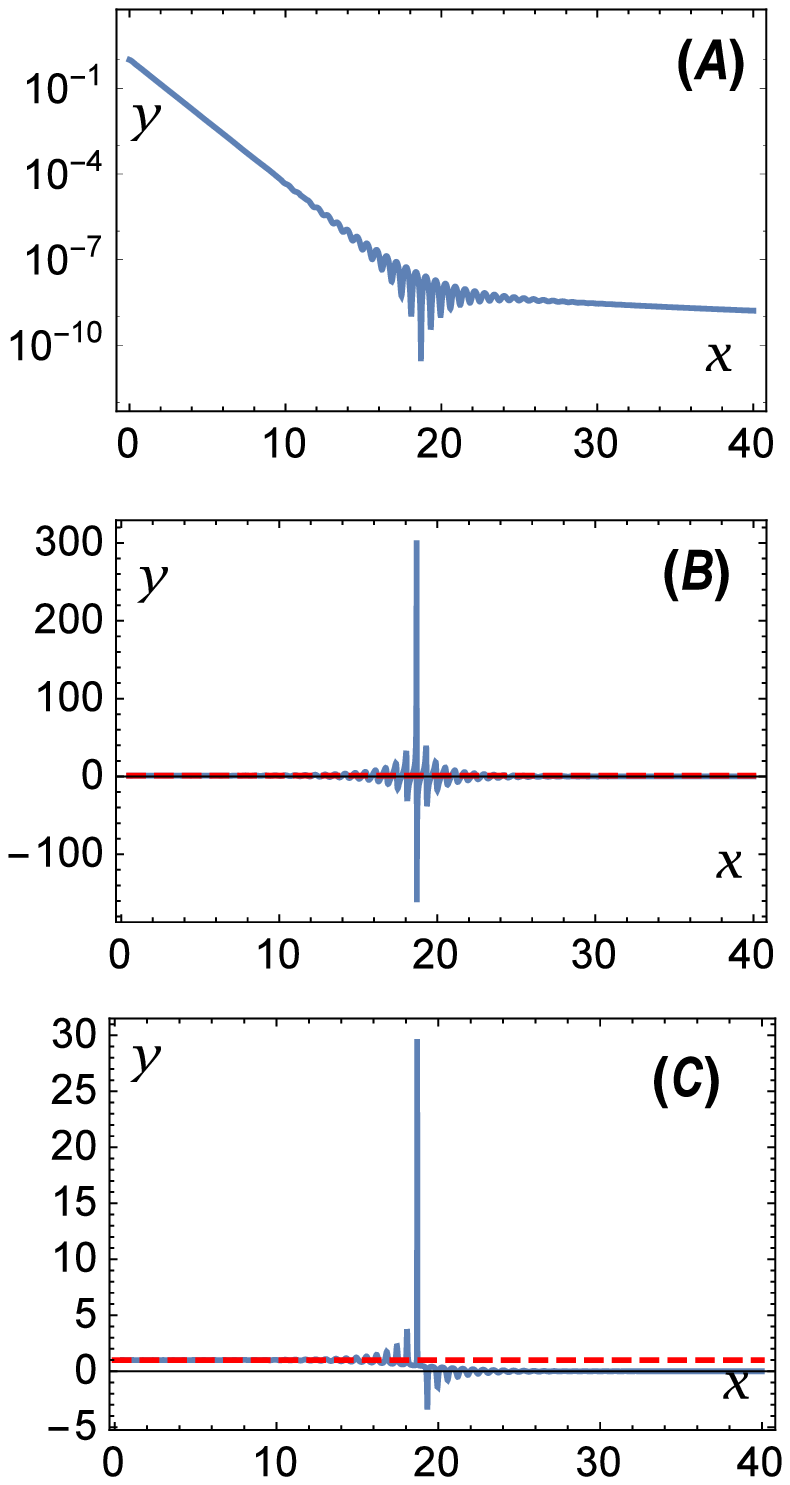}}
\caption{ $(A)$ The decay curve, $(B)$ The instantaneous decay rate, $(C)$ The instantaneous energy. Axes: $x$ --- in all panels time $t$ measured in lifetimes  $\tau$: $x = t/\tau$;  $(A)$ --- $y = {\cal P}(t/\tau)$, (The logarithmic scale); $(B)$ --- $y = \gamma (t/\tau)/ \gamma_{0}$; $(C)$ --- $\kappa (t/\tau)$, ($\kappa (t/\tau)$ is defined by formula (\ref{kappa})). The case  $\beta=10$.}
  \label{f2}
  \end{center}
  \end{figure}

\begin{figure}[H]
\begin{center}
{\includegraphics[width=78mm]{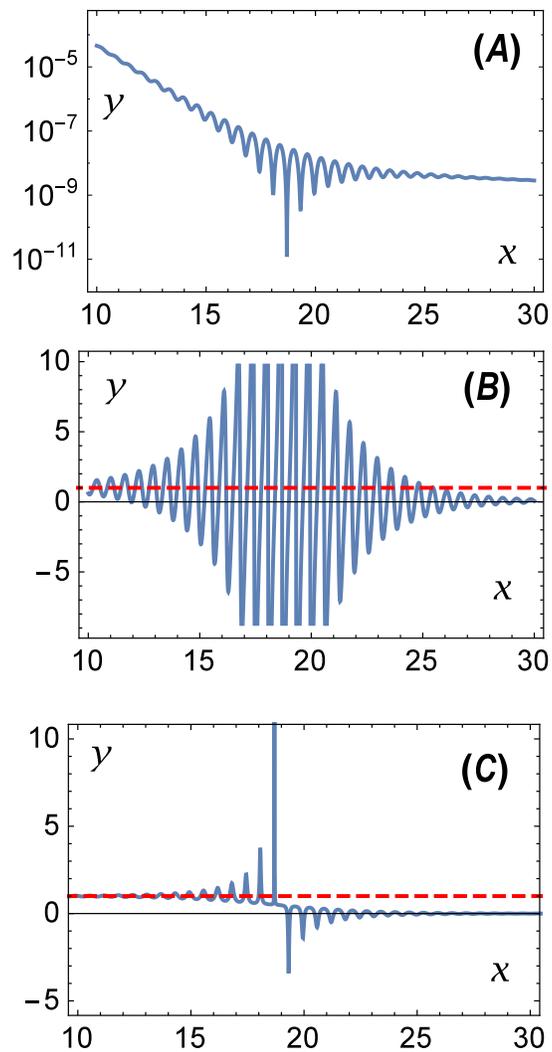}}
\caption{ The same as in the Fig (\ref{f2}): The transition times region: $t \sim T$.}
\label{f3}
  \end{center}
  \end{figure}

\begin{figure}[H]
\begin{center} %{center}
{\includegraphics[width=66mm]{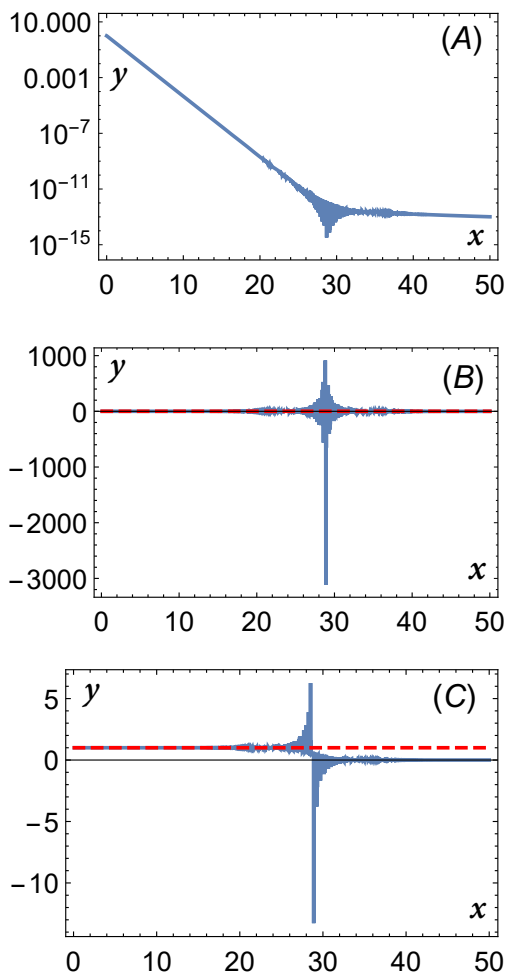}}
\caption{ $(A)$ The decay curve, $(B)$ The instantaneous decay rate, $(C)$ The instantaneous energy. Axes: $x$ --- in all panels time $t$ measured in lifetimes  $\tau$: $x = t/\tau$;  $(A)$ --- $y = {\cal P}(t/\tau)$, (The logarithmic scale); $(B)$ --- $y = \gamma (t/\tau)/ \gamma_{0}$; $(C)$ --- $\kappa (t/\tau)$, ($\kappa (t/\tau)$ is defined by formula (\ref{kappa})). The case  $\beta=100$.}
  \label{f4}
  \end{center}
  \end{figure}

This Section contains results of numerical studies of the the quantities analyzed in the previous Section.  Their results were
obtained for the chosen values of $\beta$ and
are presented  graphically
in Figs (\ref{f1}) --  (\ref {f4}). In all figures
one can find a typical form of the quantities characterizing properties of the unstable state
as a function of time: Decay curves ${\cal P}(t)$,   instantaneous decay rates $\gamma(t)$ and   instantaneous energies $E(t)$.

In all Figures: The horizontal dashed line in panels $(B)$ denotes the situation when $\frac{\gamma (t/\tau)}{\gamma_{0}} = 1$ (that is when $\gamma(t/\tau) = \gamma_{0}$) and
the horizontal dashed line in panels $(C)$ denotes
the case $\kappa (t/\tau) = 1$ which is equivalent to the condition $E(t/\tau) = E_{0}$.

\section{Discussion and final remarks}

A similar form of a decay curves to those presented in panels $(A)$ of Figs (\ref{f1}) -- (\ref{f4})
one meets for a very large class of models defined  by energy densities $\omega(E )$
of the following type
\begin{eqnarray}
\omega(E)\,&=&\, \frac{N}{2\pi}\;{\mathit \Theta}(E - E_{min} )\;(E - E_{min})^{\lambda} \, \times \nonumber \\ && \;\;\;\;\;\;\;\;\times \; \frac{\gamma_{\phi}^{0}}{(E- E_{0})^{2}+
\frac{(\gamma_{0})^{2}}{4}}\, f(E),\label{omega-gen}
\end{eqnarray}
where $\lambda \geq 0$, $f(E)$ is a form--factor ---  it is a smooth function going to zero as
${E} \rightarrow \infty$ and it has no threshold and no pole. The asymptotical large time behavior
of $a(t)$ is due to the term $(E - E_{min})^{\lambda}$ and the choice of $\lambda$. The density
$\omega(E )$ defined by the relation (\ref{omega-gen}) fulfills all physical requirements and it
leads to the decay curves having a very similar form at transition times region (where $t \sim T$)
to the decay curves presented above. The characteristic  feature of all these decay curves is the
presence of sharp and frequent oscillations at the transition times region.
This means that derivatives of the amplitude $a(t)$ may reach extremely large values for some times
from this time region and the modulus of these derivatives is much larger than the modulus of $a(t)$, which
is very small for these times. This explains why in this time region the real and imaginary parts of
$h(t) \equiv  E(t) \, - \,\frac{i}{2}\,\gamma(t)$, which can be expressed by the relation (\ref{h(t)}),
ie. by a large  derivative of $a(t)$  divided by a very small $a(t)$, reach values much larger than
the energy $ E_{0}$ of the the unstable state measured at times for which the decay curve has the exponential form.

Results presented in Sec. 3 shows that within the model considered there does not exist a finite time
interval where it could be $a(t) = a_{c}(t)$. So one can expect  $|a(t) - a_{c}(t)| \neq 0$ and varies
over time $t$ not only at times $t \sim T$ but also even for times  $t \ll T$ (see also \cite{ku-2016}).
On the other hand results of numerical calculations presented in graphical form in Sec. 4 shows that
at these times $|a(t)|^{2} \simeq |a_{c}(t)|^{2}$ to a very good approximation. So the amplitude of
the mentioned variations of $|a(t) - a_{c}(t)|$ in time for $t \ll T$ is rather almost negligible small.
Such a conclusion agrees with the conclusion resulting from analysis of the form of $E(t)$ and $\gamma (t)$
obtained numerically and presented in Sec. 4: We observe that at times $t \ll T$ $E(t) \simeq E_{0}$ and $\gamma(t) \simeq \gamma_{0}$.

Note that the equivalent formula (\ref{h(t)-eq}) for $h(t)$ means that
$h(t)$ is the so--called "weak value". The behavior of ${ E}(t)$ and $\gamma (t)$ at the transition
times region, where $ t \sim T$, presented in Figs (\ref{f1}) -- (\ref{f4}) is quite obvious for the weak values.

The question is whether and where this effect can manifest itself.
As it was mentioned earlier,
the effect presented in panels $(A)$ in Figs (\ref{f1}) -- (\ref{f4}), that is the
transition of the exponential form of the decay law ${\cal P}_{\phi}(t) = |a(t)|^{2}$   into the  inverse power law form,
was confirmed experimentally by Rhote nad his group.
This means that effects presented in panels $(B)$ and $(C)$ of these Figures   and resulting
from  the properties of the amplitude $a(t)$
have to take place for $h(t)$
too, and thus for $E(t)$ and $\gamma(t)$.

It seems that the following cases are the most likely ones where the above described
long time properties of unstable states may manifest itself or where they can be observed;
(i) One should analyze properties of unstable states having not too long values of the cross--over time $T$, or
(ii) one should find a possibility to observe a suitably large number of events, i.e. unstable particles, created by the same source.

The problem of broad resonances in the scalar sector  ($\sigma$ meson problem) discussed in \cite{nowakowski1,nowakowski2},
where the hypothesis was formulated that this problem could be connected with properties of the decay amplitude in the transition time region,
seems to be  possible manifestations of this effect and this problem refers  to the first possibility mentioned above.

Astrophysical and cosmological processes in which  extremely huge numbers of unstable particles are created
seem to be  another possibility for the above discussed effect to become manifest. The probability
${\cal P}_{\phi}(t) = |a(t)|^{2}$ that an unstable particle, say $\phi$, survives up to time $t \sim T$ is extremely small:
 Let ${\cal P}_{\phi}(t)$ be such that
\begin{equation}
{ {\cal P}_{\phi}(t)\,\vline }_{\;t \sim T}\;\sim\;10^{-k},
\label{p(t)-k}
\end{equation}
where $k \gg 1$, then there is a chance to observe some of particles $\phi$ survived at
$t \sim T$ only if there is a source creating these particles in ${\cal N}_{\phi}$ number such that
\begin{equation}
{{\cal P}_{\phi}(t)\,\vline }_{\;t \sim T}\;{\cal N}_{\phi} \; \gg \;1.
\label{N-phi}
\end{equation}
So if  a source exists that creates a flux containing
\begin{equation}
{\cal N}_{\phi} \;\sim\;10^{\,l},
\label{N-phi-l}
\end{equation}
unstable particles and $l \gg k$ then
the probability theory states that the number $N_{surv}$ unstable particles\
\begin{equation}
N_{surv} = { {\cal P}_{\phi}(t)\,\vline }_{\;t \sim T}\;{\cal N}_{\phi} \;\sim\;10^{l - k} \; \gg\;0,
\label{N-surv}
\end{equation}
has to survive up to time $t \sim T$.

Sources creating such numbers of unstable particles are known from cosmology and astrophysics:
The Big Bang;
 Processes taking place in  galactic  nuclei (galactic cores);
Processes taking place inside  stars (supernova explosions);
Etc..
So one should look for a manifestation of the quantum effect presented in panels $(B)$
and $(C)$ in Figs (\ref{f2}) -- (\ref{f4}) in astronomical observations and in cosmology.

The third case, where late time properties of the instantaneous energy $E(t)$ of
the unstable state can can play important role is cosmology.
It seems also that the long time properties of the energy $ E(t)$ of unstable states
can justify the use time dependent cosmological constant $\Lambda$
within the so--called $\Lambda$ Cold Dark Matter ($\Lambda$CDM) Cosmology.
 Namely form the literature it is known that
 the cosmological  constant $\Lambda$ of the form,
\begin{equation}
\Lambda \equiv \Lambda (t) \simeq \Lambda_{bare} + \frac{D}{t^{2}}, \label{Lambda-t}
\end{equation}
(where $D = const.$) was considered in many papers:
Similar form of $\Lambda$ was obtained in \cite{Canuto}, where
the  invariance under scale transformations of
the generalized Einstein equations was studied.
Such a time dependence of $\Lambda$ was
postulated also in \cite{Lau} as the result of
the analysis of the large numbers hypothesis.
The cosmological model with  time dependent {$\Lambda$}
of the above postulated form was studied also in \cite{Berman}.
This form of $\Lambda$  was assumed in eg.
in \cite{Lopez} but there was no any explanation what
physics suggests such the choice. Cosmological model with
time dependent {$\Lambda$} were also studied in much more recent papers.

Krauss and Dent in their paper \cite{krauss1,krauss2}
made a hypothesis
that some false vacuum regions do survive well up to the time $T$ or  later.
So, let $|\phi\rangle = | 0\rangle^{false}$,
be a false, $|0\rangle^{bare}$  -- a bare, true vacuum states and  ${E}_{0} \equiv E^{false}_{0}$
 be the energy of a state corresponding to the false vacuum measured at the canonical decay time
and  $E^{bare}_{0} \equiv { E}_{min}$ be the energy of true vacuum (i.e. the true ground state of the system).
As it is seen from the results presented in  Section 3, the problem is that the energy of those false
vacuum regions which survived up to $T$ and much later differs from $E^{false}_{0}$.
Going from quantum mechanics to quantum field theory one should take into account
among others a volume factors so that survival probabilities per unit volume per
unit time should be considered. The standard false vacuum decay calculations shows
that the same volume factors should appear in both early and late time decay rate
estimations (see Krauss and Dent \cite{krauss1}). This means that the calculations
of cross--over time $T$ can be applied to survival probabilities per unit
volume.  For the same reasons  within the quantum field theory the quantity
${ E}(t)$ can be replaced by  the energy per unit volume  $\rho(t) = { E}(t)/V$
because these volume factors $V$  appear in the numerator and denominator of the
formula (\ref{h(t)}) for $h(t)$.
Therefore assuming that we live in the Unverse with a false vacuum and  based
on the result (\ref{Re-h-as}) one concludes that there should be at times $t > T$:
\begin{equation}
{ E}^{false}(t) \simeq E_{bare}^{0}  + \frac{c_{2}}{t^{2}} + \frac{c_{4}}{t^{4}}\ldots ,\;\;\; {\rm for}\;\;\; t \gg T,
\label{E-false-infty-1}
\end{equation}
(where $c_{2} = c_{2}^{\ast}$ and $c_{4} = c_{4}^{\ast}$),
or,
\begin{equation}
\rho_{0}(t)  \stackrel{\rm def}{=} {\rho}^{false}(t) \simeq \rho^{bare}_{0}  + \frac{d_{2}}{t^{2}} + \frac{d_{4}}{t^{4}}\ldots ,\;\;\; {\rm for}\;\;\; t \gg T,
\label{rho-false-infty-2}
\end{equation}
(where $\rho^{false}(t)$ is the energy density in the false vacuum state,
$d_{2}=d_{2}^{\ast},\;d_{4}=d_{4}^{\ast}$, ${\rho}_{0}(t)= {\rho}^{false}(t) ={ E}^{false}(t)/V $, $ \rho_{0}^{bare}=  E^{bare}_{0}/V$).
The standard relation is
\begin{equation}
\rho_{0} \equiv \rho^{bare}_{0} = c^{2}\, \frac{\Lambda_{0}}{8 \pi G}, \label{Lambda-rho}
\end{equation}
where $\Lambda_{0} \equiv \Lambda^{bare}$ is the bare cosmological constant. \\

So the relations
\begin{equation}
\rho_{0} (t) = \rho_{0}^{bare} + \frac{d_{2}}{t^{2}} + \frac{d_{4}}{t^{4}}, \;\;\;{\rm and}\;\;\;\;\;
\Lambda (t) = \Lambda^{bare} + \frac{D_{2}}{t^{2}} + \frac{D_{4}}{t^{4}}, \label{rho-Lambda}
\end{equation}
(where $D_{2}, D_{4}$ are real) are equivalent and they both are a manifestation of quantum long
time properties of  unstable states. These last two relations explain why
it is reasonable to use $\Lambda$ of the form (\ref{Lambda-t}), especially when one considers cosmologies with the false vacua.

Summing up: Late time properties of evolving in time quantum unstable systems are
extremely difficult to observe. Discussion presented in this Section shows where there
is a chance to observe a manifestation of these late time effects and in which cases
taking into account these properties may explain why some theories are worth to a deeper
analysis: A good example is the cosmology with time dependent cosmological "constant" $\Lambda (t)$.

\end{document}